# Explorations in Designing Virtual Environments for Remote Counselling

Jiashuo Cao*
The University of Auckland

Wujie Gao†
The Hong Kong Polytechnic University

Yun Suen Pai‡
The University of Auckland

Simon Hoermann§
University of Canterbury

Chen Li¶
The Hong Kong Polytechnic University

Nilufar Baghaei‖
The University of Queensland

Mark Billinghurst**
The University of Auckland

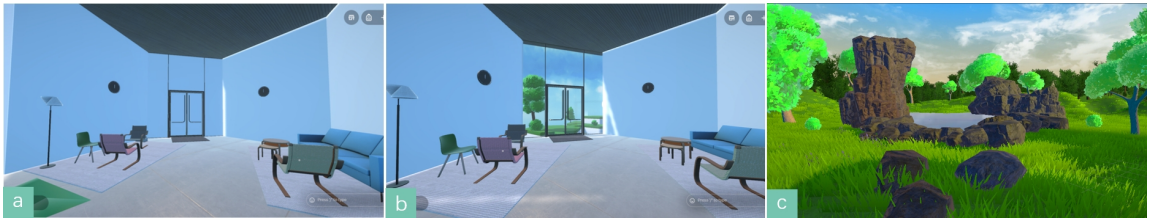

Figure 1: Screenshots of the VR test environments. a: enclosed environment. b: semi-open environment. c: open environment

**ABSTRACT**

The advent of technology-enhanced interventions has significantly transformed mental health services, offering new opportunities for delivering psychotherapy, particularly in remote settings. This paper reports on a pilot study exploring the use of Virtual Reality (VR) as a medium for remote counselling. The study involved four experienced psychotherapists who evaluated three different virtual environments designed to support remote counselling. Through thematic analysis of interviews and feedback, we identified key factors that could be critical for designing effective virtual environments for counselling. These include the creation of clear boundaries, customization to meet specific therapeutic needs, and the importance of aligning the environment with various therapeutic approaches. Our findings suggest that VR can enhance the sense of presence and engagement in remote therapy, potentially improving the therapeutic relationship. In the paper we also outline areas for future research based on these pilot study results.

## 1 INTRODUCTION

The landscape of mental health services has experienced a significant transformation with the advent of technology-enhanced interventions [10]. Traditional face-to-face psychotherapy, while effective, poses accessibility challenges for individuals in remote areas or those unable to attend in-person sessions due to various barriers such as geographical constraints, physical disabilities, or social stigma [26]. Telehealth solutions, particularly video conferencing, have emerged as a viable alternative, offering the potential to bridge these gaps [11]. However, the perceived remoteness and lack of presence in such treatments may impede the therapeutic process, limiting the depth of client engagement [15, 27].

*e-mail: jcao403@aucklanduni.ac.nz
†e-mail: wujiegao@polyu.edu.hk
‡e-mail: yun.suen.pai@auckland.ac.nz
§e-mail: simon.hoermann@canterbury.ac.nz
¶e-mail: richard-chen.li@polyu.edu.hk
‖e-mail: n.baghaei@uq.edu.au
**e-mail: mark.billinghurst@auckland.ac.nz

Recent advancements in Virtual Reality (VR) technology offer a promising avenue for enhancing remote psychotherapy by creating a sense of presence and immersion that closely mimics in-person interactions [25, 29]. VR's ability to simulate realistic environments and social interactions through avatars could potentially overcome some of the limitations associated with traditional video conferencing methods [11]. Moreover, the immersive nature of VR might facilitate a deeper level of engagement and rapport between therapists and clients, essential elements for effective psychotherapy [19]. VR also enables the creation of immersive environments specifically designed to create therapeutic experiences, such as conducting art therapy through spatial painting [16], or performing compassion focused therapy[1].

However, while there is growing evidence supporting the use of VR in therapy, much of the existing research has primarily focused on single user exposure therapy, using the immersive nature of VR [4, 7]. There is considerably less research on the application of VR for remote counselling, which emphasizes the co-presence of therapists and clients in a virtual environment. Furthermore, while the environment plays a crucial role in face-to-face therapy [28], there is limited research on how the virtual environment impacts the effectiveness of remote therapy delivered via VR. Understanding how elements like virtual surroundings affect therapeutic outcomes could be essential for optimizing the efficacy of VR-based counseling.

This work presents a pilot study investigating some key factors in designing virtual environments for remote counselling. We aim to identify some essential elements that could facilitate a sense of presence, engagement, and therapeutic alliance between clients and therapists. By conducting a series of interviews and evaluations in different VR environments with experienced psychotherapists, we explored how these virtual spaces could be tailored to support remote counselling and enhance the overall counselling experience. Our findings provide valuable insights into the potential of VR to revolutionize remote mental health services and highlight the importance of thoughtful design in creating effective therapeutic environments. The main novelty of this research is that it collects feedback from practicing therapists using a range of different VR spaces, and then performs a thematic analysis on interviews with the therapists to identify the key factors mentioned.

## 2 RELATED WORK

### 2.1 Remote Therapy Using Non-VR Technology

Remote therapy is traditionally done via videoconferencing, telephone, or messaging interventions, following the structure of a typical face to face psychotherapy session. However, these each of these approaches can encounter communication limitations [20]. For example, due to lack of spatial cues, videoconferencing may not fully replicate the sense of presence or feeling of sharing the same physical space as with a face to face therapist [31]. Despite this, videoconferencing has been shown to be effective in delivering psychological treatments [24]. Similarly, telephone therapy lacks the transmission of critical visual and non-verbal cues, which is essential in conventional therapy settings. Nonetheless, studies by Mohr et al. [21] acknowledge its potential in mental health care, suggesting its comparability to face-to-face sessions. Finally, Messaging cannot convey vocal communication cues, but is also validated by research for its effectiveness in psychological treatments [12]. Overall, this research shows that there are limitations with current telehealth solutions, but despite this, they can contribute to mental health support.

### 2.2 Remote Therapy Using VR

Some of the limitations of current remote therapy solutions could be overcome through the use of Virtual Reality (VR). Using VR for therapy has been explored through pioneering work, notably in treating anxiety disorders [5, 22], phobias [14], and post-traumatic stress disorder (PTSD) [8]. Early studies demonstrated the efficacy of VR in providing exposure therapy in a controlled, immersive environment, facilitating significant progress in understanding and treating complex mental health conditions[13]. These foundational efforts laid the groundwork for the development of VR applications tailored to a wide range of therapeutic needs, emphasizing the potential of VR to revolutionize the field of psychotherapy by offering personalized, engaging, and effective treatment modalities [4].

While most use of VR for therapy has focused on single user experiences, there are some researchers who have explored how shared VR experience could support psychotherapy. For example, Pedram et al.[23] introduced VRchat as a remote counselling tool, where therapist and client experience a shared virtual environment. They found that counselling through VR outperformed using a telephone in terms of communicate effectiveness, realism, and the sense of presence experienced by clients and therapists. However, they focused on participants' feelings about the interview process and did not assess the actual therapeutic effects, or evaluate the VR environment used. Similarly in the research from Matsangidou et al.[19], they explored the use of a Multi-User VR (MUVR) system as a therapeutic medium for individuals at high risk for developing eating disorders. While they demonstrated that MUVRs can effectively enhance therapeutic practices like Acceptance and Commitment Therapy, Play Therapy, and Exposure Therapy, the discussion of how the virtual environment could support therapy is missing.

In summary, early research has established the effectiveness of VR in providing controlled, immersive environments for psychotherapy, paving the way for VR applications tailored to various therapeutic needs. While much of the existing work has focused on single-user experiences, a few studies have examined the potential of shared VR environments in psychotherapy. However evaluation of the design of the virtual environment is usually ignored. Thus, our work focus on uncovering the benefits of the virtual environment in VR-based remote counselling, by gathering feedback from experienced psychotherapists on the design of therapeutic VR environments.

## 3 RESEARCH METHODS

In our research we were interested in exploring what features of virtual environments would make them most suitable for remote counselling. We conducted a pilot study by having participants experience three different virtual environments and interviewing them. Through this process, we aimed to answer the following research question: *What factors should be considered when designing an immersive virtual environment for remote counselling?* . In this section, we will provide more details about the participants, how we designed and developed the virtual environments, the evaluation process, and the data analysis.

### 3.1 Participants

Our participants were four experienced psychotherapists (P1-P4), who had been working in the field for 3 to 15 years each with an average of 7 years. Three of the therapists (P2 - P4) mainly provide services to adults, while P1 primarily works with teenagers and college students. The therapeutic approaches they commonly use include Cognitive Behavioral Therapy (CBT) [3](P1,P2,P4), Interpersonal Therapy (IPT) [17](P1, P3), Acceptance and Commitment Therapy (ACT) [9] (P2), Dialectical Behavior Therapy (DBT) [18] (P2), and Drama Therapy [6] (P4). All therapists had experience conducting remote counselling sessions. In terms of head-mounted VR usage, only P1 has had a few experiences with Meta Quest 2, while the other therapists had not used VR head-mounted displays (HMDs) before. Table 1 provides more detail about the participants in the pilot study.

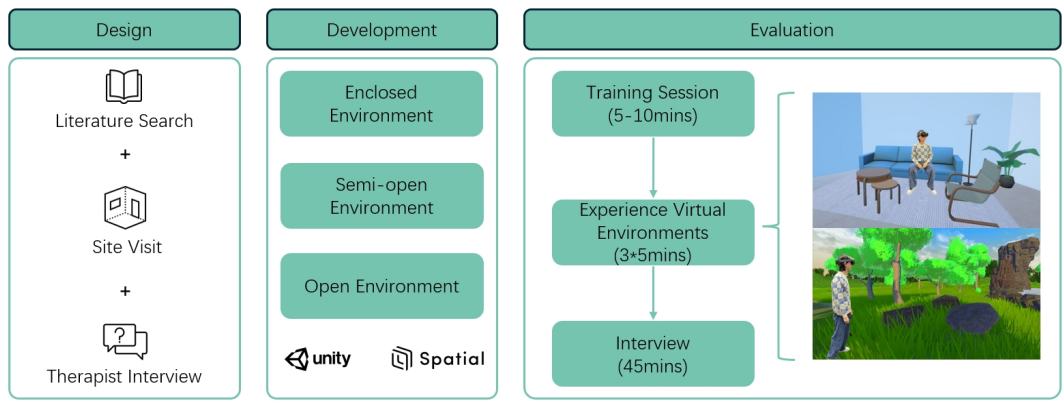

Figure 2: Design, development and evaluation process

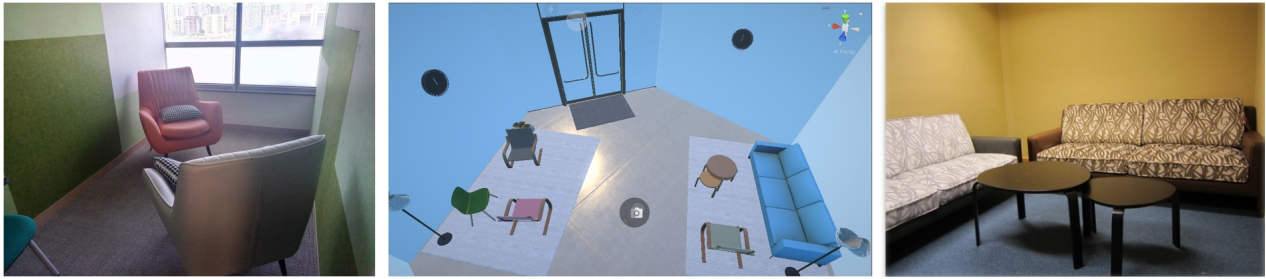

Figure 3: Left, Right: Pictures of therapy rooms. Middle: a VR environment combining two types of rooms.

| ID | Gender | Years of Counselling | Client group | Therapy Type |
|----|--------|---------------------|--------------|--------------|
| P1 | Female | 3 | Adolescents | CBT, IPT |
| P2 | Male | 15 | Adults | CBT, DBT, ACT |
| P3 | Female | 6 | Adults | IPT |
| P4 | Male | 4 | Adults | CBT, Drama Therapy |

Table 1: Overview of the psychotherapists

### 3.2 Virtual Environments

**Design:** The design of the virtual environments used in this experiment was informed by three sources: previous related research, real-life counselling rooms, and interviews with therapists. First, we collected common types of virtual environments frequently mentioned in prior research, including enclosed spaces (e.g.[29]), semi-open spaces (e.g. [30]), and open spaces (e.g. [19]). We then took photographs of counselling rooms at the university and drew floor plans showing the layout of these rooms. Finally, we conducted an interview with P1, focusing on what environmental elements they considered important in a counselling room. During the interview, we used the previously taken photographs and floor plans to help clarify the therapist's opinions. Based on all the collected data, we established the following design principles: seating arrangements should allow clients to avoid direct eye contact with the therapist; the environment should include an object to help therapists manage time; appropriate lighting and colours should create a relaxing atmosphere; and the setup should ensure clients feel that their privacy is protected.

**System Development** Based on our design principles, we proceeded to develop three test VR environments: (1) an enclosed environment with two different seat settings, several indoor plants and clocks on the wall (see fig3), (2) a semi-open space which has same room setting as the enclosed environment but with a large window in the room that offers a view of the natural scenery outside, and (3) an open environment with three stones serving as seats in a forest with a pond opposite them (see fig1). Development was carried out using the Unity platform, enabling the creation of immersive and interactive 3D environments. These spaces were then published on the Spatial.io[1] social VR platform, ensuring accessibility through both web browsers and VR headsets for a wide range of users. Spatial.io also supports collaborative viewing, allowing multiple users to interact and explore the virtual environment together in real time.

### 3.3 The Evaluation Process

We used the Meta Quest Pro as the VR HMD in our evaluation. Before the evaluation, we provided each therapist with a training session lasting 5 to 10 minutes to help them learn some basic skills in VR, such as how to use the handheld controller to move in the environment. After the training session, therapists were asked to first enter and observe the virtual environment. They were then instructed to imagine conducting counselling sessions with their typical client group in this environment and to provide feedback and opinions based on this context. Some therapists requested to have another avatar in the same environment to create the feeling of counselling. During this process, researchers guided the therapists by asking questions to elicit their thoughts, some questions focused on therapists feeling of the environment, e.g. *How did you feel about the overall atmosphere of the virtual environment?*. Other questions asked about their ideas on using the virtual environment to conduct counselling, e.g. *Do you think this environment is suitable for counselling? Why?*.

---

[1] https://www.spatial.io/

After experiencing the three virtual environments, the therapists participated in a semi-structured interview. The primary aim of the interview was to understand their overall experience with the virtual environments and their views on how these environments could support remote counselling. We asked questions such as *What did you like most about the virtual environment?* and *Are there any specific elements or features you believe should be added to enhance the therapeutic experience in VR?*. The entire evaluation process took approximately one hour, of which participants typically spent 5 minutes in each VR environment (see fig2). The interview was recorded for analysis later.

### 3.4 Data Analysis

We gathered a total of 245 minutes of interview audio data, which were subsequently transcribed into text using Microsoft Word's transcription function and subsequently reviewed and verified for accuracy by a researcher. We then conducted a thematic analysis on these transcripts, following the general guidelines established by Braun and Clarke [2]. This analysis was a collaborative effort between two researchers: a developer and a psychotherapist. Initially, all transcripts were thoroughly reviewed, and one researcher assigned codes to the relevant data, organizing these into overarching themes. We used 37 codes in three categories: User experience, e.g., codes of immersion, presence, and relaxation; Therapeutic factors, e.g., codes of privacy, safety, and therapeutic approaches; General feedback, e.g. codes of likes, dislikes, and suggestions. Another researcher then reviewed these themes, codes and accompanying comments, ultimately refining and defining the final key themes. We present the result based on the final key themes.

## 4 PILOT STUDY RESULTS

After analyzing the data, we identified two main categories of key themes: one category consisted of themes strongly associated with the specific environments shown, while the other included general themes related to virtual environments as a whole. We will use this categorization as the structure to report our main themes in the rest of this section.

## 4.1 Environment-related Theme

**Enclosed Environment** All therapists appreciated the calming colours and ambiance of the enclosed environment, noting that it provided a comforting space for therapy sessions. For example, participant P2 said, *"I really like the colours in here, they make me feel calm and relaxed."*. However, the inclusion of two sets of seating arrangements raised concerns. One therapist suggested that this configuration might induce anxiety in clients, as it could imply the possibility of another person entering the space. *"If there are two sets of seating in the room, someone undergoing therapy might wonder, 'If I sit here, could someone else come in on the other side?' This could make the room feel less private and reduce the sense of security for the person."*(P3). To enhance focus and provide a sense of security during counseling, one therapist recommended opting for a smaller, more enclosed space. *"Small, enclosed spaces can foster a sense of security while helping the client focus on counselling, which is really helpful during the therapy."*(P3).

**Semi-Open Environment** The semi-open environment received positive feedback for that bright external lighting and the openness that it provided. After experiencing the space, P1 highlighted the psychological benefits of maintaining a connection with the external environment, such as the uplifting effects of sunlight and expansive views. P1 said, *"The main benefit of semi-open spaces is that the external environment can help maintain a positive mindset for users, such as feeling the sunshine or viewing expansive scenes, which are helpful."*. Notably, therapists appreciated the scenery in semi-open environment for its 'virtual feel'. As P4 described, *"I find the scenery outside this room quite well-designed because it's not something you'd typically encounter in real life. This gives clients a subconscious reassurance that there won't be any passersby or potential onlookers outside."*. Therapists also pointed out that in a semi-open room, clients should be able to see the external environment from any seating position. *"You might consider putting a window between the two chairs, so that no matter where you sit, you can see outside."*(P4).

**Open Environment** All therapists expressed their appreciation for the environment on their first entry into the open space. For example, P2 said *"The grass and pond setting is very comfortable and makes it easy to relax."*. The natural environment was described as refreshing and conducive to relaxation, but P1 also expressed concerns about its suitability for individuals with attention difficulties, such as children with ADHD. The richness of content in this environment, including elements like trees and lakes, could potentially distract clients. *"The forest (open space) does make me feel comfortable... But I'm not sure if clients might easily get distracted staying in this environment; they might wonder what's beyond that lake."*(P1). Similar concerns were mentioned by P3, who noted that the numerous visual stimuli and vibrant colours might make it challenging for clients to focus on the therapy session. This could potentially leading to decreased effectiveness of the therapeutic process. *"I'm not confident I can keep the clients focused in this environment. The colours here are a bit too bright, and the trees and grass are constantly moving in the wind."* (P3).

## 4.2 Shared Themes About the Virtual Environment

**Creating Boundaries in Virtual Environment to Facilitate Therapeutic Alliance** One of the significant findings from the study is the importance of creating clear boundaries within the virtual environment to support the establishment of therapeutic alliances. Most therapists mentioned it when they were asked to imagine that they needed to conduct counselling in the open environment. For example, P2 said *"This setting(open environment) feels too open to me. I'd rather do counselling in previous rooms because it makes me feel like I'm in the same space with them, which makes it easier to connect with them."*. P4 emphasized that creating boundaries is not limited to using walls for visual separation; other objects can also provide a sense of boundaries, such as carpets and pathways. *"It's not necessary to enclose an area with walls to establish boundaries. For instance, the carpet you placed in the seating area within the room creates a boundary... Outdoors, you could use a small path to encircle those stones, and that also forms a specific area, which I could use in the counselling"*. Therapists also highlighted that defined boundaries help both therapists and clients understand the scope and limits of the session, and the physical boundaries present in a traditional counselling setting could help influence the perception and maintenance of these boundaries. This clarity was found to foster a sense of safety and professionalism, which are critical for building trust and a strong therapeutic relationship.

**Customisability of the Virtual Environment but limited** The second theme centers on being able to customize the virtual environment. Participants emphasized the need for the virtual space to be adaptable to cater to their specific needs and preferences. P1 expressed a desire to replace the room's clock with a digital clock, stating, *"I'm more used to reading digital clocks; it can just be placed on the table instead of hanging on the wall."*. P2 wanted to have a writable whiteboard in the room to assist with performing CBT. P3 hoped to be able to adjust the seating arrangement. P4 suggested the ability to adjust the lighting in open environment to help clients understand the remaining time, saying, *"I think it would be great if I could change the position of the sun, for example, setting it to a sunset state when there are ten minutes left in the session."*. However, they all mentioned that they don't want to build an environment entirely from scratch, as they believe it to be time-consuming and unnecessary. They assume it's sufficient to provide customization options for specific elements or features instead.

**Tailoring Environments to Different Therapeutic Approaches** The third theme identified the necessity of tailoring virtual environments to support various therapeutic approaches. Participants noted that different types of therapies might require distinct environmental setups to be effective. For instance, a minimalist and distraction-free setting might be preferable for CBT, while a more relaxed and open environment could be beneficial for therapies focusing on interpersonal dynamics. P4 said *"If I'm doing CBT, I would definitely use one of the first two environments(enclosed environment and semi-open environment). But if it's art therapy, the third environment might be better. For drama therapy, I'd want the environment to match the setting needed for each role-playing scenario."*. This adaptability allows therapists to align the virtual environment with their therapeutic goals and methods, thus optimizing the effectiveness of the treatment. In addition, ensuring that the furniture sets matched was advised to avoid conveying unintended messages during therapy sessions. *"Typically, we only set up an extra chair in the room when we need to do an empty chair exercise. So, the room's arrangement depends on the therapy planned for that day."*(P1).

## 4.3 Limitations

Although these results are very interesting, there are a number of limitations with this work that will have to be addressed in the future. First, this was a pilot study with a limited number of psychotherapists who each explored the space for a short time. In the future, we would like to complete a full study with a sufficient number of psychotherapists to perform a comprehensive statistical analysis of the results. The study also did not include people role-playing clients within the virtual environments, which limits the ability to fully assess the practical implications of using these spaces for therapy sessions. The presence of clients would provide more accurate insights into the therapeutic dynamics and effective-ness of the virtual setting.

Another limitation is the restricted scope of experimental measures. The study relied primarily on a few subjective questions and results, lacking the use of validated surveys that could provide more robust data. Future studies should incorporate standardized measures such as surveys on presence, the System Usability Scale (SUS), and other relevant metrics to better evaluate the user experience and system performance.

Finally, the study only used a limited number of VR spaces with limited interactions. Exploring a broader range of virtual environments with diverse features would be valuable in understanding how different aspects of virtual design impact the therapeutic process. This exploration could reveal specific elements that enhance or hinder the therapeutic experience, guiding more effective design strategies for virtual counselling environments.

## 5 Conclusions and Future Work

The use of VR for remote counselling presents a transformative potential for enhancing mental health services, particularly in overcoming the limitations of traditional remote therapy tools (video conferencing, phone calls). This pilot study explored the key factors in designing virtual environments that are conducive to effective therapy sessions, focusing on elements such as the sense of presence, engagement, and the quality of the therapeutic relationship. Our findings underscore the significance of creating well-defined boundaries within virtual spaces to foster a therapeutic alliance, addressing the need for customization to suit various therapeutic approaches, and ensuring the environment can be tailored to individual needs.

In the future, we plan to gather input from a larger and more diverse group of therapists to conduct a more comprehensive analysis of their thoughts on the effectiveness and usability of virtual environments for remote counselling. To gain deeper insights, we intend to involve therapists in role-playing sessions within the virtual environments where they can take on different roles such as the therapist and the client. We will them collect valuable feedback on the dynamics and interactions between multiple people within the virtual space. Workshops are also getting organized aimed at evaluating and analyzing how other elements in VR can support remote counselling, such as different avatar representations and interaction methods. By understanding how these components can be optimized, we aim to create more engaging and supportive virtual environments for both therapists and clients.

## References

[1] N. Baghaei, L. Stemmet, I. Khaliq, A. Ahmadi, I. Halim, H.-N. Liang, W. Xu, M. Billinghurst, and R. Porter. Designing individualised virtual reality applications for supporting depression: A feasibility study. In *Companion of the 2021 ACM SIGCHI symposium on engineering interactive computing systems*, pp. 6–11, 2021. 1

[2] V. Braun and V. Clarke. Using thematic analysis in psychology. *Qualitative research in psychology*, 3(2):77–101, 2006. 3

[3] A. C. Butler, J. E. Chapman, E. M. Forman, and A. T. Beck. The empirical status of cognitive-behavioral therapy: A review of meta-analyses. *Clinical psychology review*, 26(1):17–31, 2006. 2

[4] P. M. Emmelkamp and K. Meyerbröker. Virtual reality therapy in mental health. *Annual review of clinical psychology*, 17:495–519, 2021. 1, 2

[5] P. M. Emmelkamp, K. Meyerbröker, and N. Morina. Virtual reality therapy in social anxiety disorder. *Current psychiatry reports*, 22:1–9, 2020. 2

[6] R. Emunah. *Acting for real: Drama therapy process, technique, and performance*. Routledge, 2019. 2

[7] F. Ferraioli, L. Culicetto, L. Cecchetti, A. Falzone, F. Tomaiuolo, A. Quartarone, and C. M. Vicario. Virtual reality exposure therapy for treating fear of contamination disorders: A systematic review of healthy and clinical populations. *Brain Sciences*, 14(5):510, 2024. 1

[8] E. B. Foa, T. M. Keane, M. J. Friedman, and J. A. Cohen. *Effective treatments for PTSD: practice guidelines from the International Society for Traumatic Stress Studies*. Guilford Press, 2010. 2

[9] S. C. Hayes and H. Pierson. *Acceptance and commitment therapy*. Springer, 2005. 2

[10] H. Herrman, C. Kieling, P. McGorry, R. Horton, J. Sargent, and V. Patel. Reducing the global burden of depression: a lancet–world psychiatric association commission. *The Lancet*, 393(10189):e42–e43, 2019. 1

[11] E. Humer, P. Stippl, C. Pieh, R. Pryss, T. Probst, et al. Experiences of psychotherapists with remote psychotherapy during the covid-19 pandemic: cross-sectional web-based survey study. *Journal of medical Internet research*, 22(11):e20246, 2020. 1

[12] D. Kessler, G. Lewis, S. Kaur, N. Wiles, M. King, S. Weich, D. J. Sharp, R. Araya, S. Hollinghurst, and T. J. Peters. Therapist-delivered internet psychotherapy for depression in primary care: a randomised controlled trial. *The Lancet*, 374(9690):628–634, 2009. 2

[13] M. Krijn, P. M. Emmelkamp, R. P. Olafsson, and R. Biemond. Virtual reality exposure therapy of anxiety disorders: A review. *Clinical psychology review*, 24(3):259–281, 2004. 2

[14] M. Krijn, P. M. Emmelkamp, R. P. Ólafsson, M. Bouwman, L. J. Van Gerwen, P. Spinhoven, M. J. Schuemie, and C. A. Van der Mast. Fear of flying treatment methods: virtual reality exposure vs. cognitive behavioral therapy. *Aviation, space, and environmental medicine*, 78(2):121–128, 2007. 2

[15] D. Lester. The use of the internet for counseling the suicidal individual: Possibilities and drawbacks. *OMEGA-Journal of Death and Dying*, 58(3):233–250, 2009. 1

[16] C. Li and P. Y. Yip. Remote arts therapy in collaborative virtual environment: A pilot case study. *Frontiers in Virtual Reality*, 4:1059278, 2023. 1

[17] J. D. Lipsitz and J. C. Markowitz. Mechanisms of change in interpersonal therapy (ipt). *Clinical psychology review*, 33(8):1134–1147, 2013. 2

[18] T. R. Lynch, A. L. Chapman, M. Z. Rosenthal, J. R. Kuo, and M. M. Linehan. Mechanisms of change in dialectical behavior therapy: Theoretical and empirical observations. *Journal of clinical psychology*, 62(4):459–480, 2006. 2

[19] M. Matsangidou, B. Otkhmezuri, C. S. Ang, M. Avraamides, G. Riva, A. Gaggioli, D. Iosif, and M. Karekla. "now i can see me" designing a multi-user virtual reality remote psychotherapy for body weight and shape concerns. *Human–Computer Interaction*, 37(4):314–340, 2022. 1, 2, 3

[20] D. C. Mohr, P. Cuijpers, and K. Lehman. Supportive accountability: a model for providing human support to enhance adherence to ehealth interventions. *Journal of medical Internet research*, 13(1):e30, 2011. 2

[21] D. C. Mohr, S. L. Hart, L. Julian, C. Catledge, L. Honos-Webb, L. Vella, and E. T. Tasch. Telephone-administered psychotherapy for depression. *Archives of general psychiatry*, 62(9):1007–1014, 2005. 2

[22] M. V. Navarro-Haro, M. Modrego-Alarcon, H. G. Hoffman, A. Lopez-Montoyo, M. Navarro-Gil, J. Montero-Marin, A. Garcia-Palacios, L. Borao, and J. Garcia-Campayo. Evaluation of a mindfulness-based intervention with and without virtual reality dialectical behavior therapy® mindfulness skills training for the treatment of generalized anxiety disorder in primary care: a pilot study. *Frontiers in psychology*, 10:414878, 2019. 2

[23] S. Pedram, S. Palmisano, P. Perez, R. Mursic, and M. Farrelly. Examining the potential of virtual reality to deliver remote rehabilitation. *Computers in Human Behavior*, 105:106223, 2020. 2

[24] L. K. Richardson, B. C. Frueh, A. L. Grubaugh, L. Egede, and J. D. Elhai. Current directions in videoconferencing tele-mental health research. *Clinical Psychology: Science and Practice*, 16(3):323, 2009. 2

[25] D. P. Rowland, L. M. Casey, A. Ganapathy, M. Cassimatis, and B. A. Clough. A decade in review: A systematic review of virtual reality interventions for emotional disorders. *Psychosocial Intervention*, 31(1):1, 2022. 1

[26] B. Saraceno, M. van Ommeren, R. Batniji, A. Cohen, O. Gureje,

J. Mahoney, D. Sridhar, and C. Underhill. Barriers to improvement of mental health services in low-income and middle-income countries. *The Lancet*, 370(9593):1164–1174, 2007. 1

[27] R. C. Scharff and V. Dusek. *Philosophy of technology: The technological condition: An anthology*. John Wiley & Sons, 2013. 1

[28] M. L. Siegel, E. M. G. Binder, H.-S. J. Dahl, N. Czajkowski, K. Critchfield, P. Høglend, and R. Ulberg. Therapeutic atmosphere in psychotherapy sessions. *International Journal of Environmental Research and Public Health*, 17, 2020. doi: 10.3390/ijerph17114105 1

[29] M. Slater, S. Neyret, T. Johnston, G. Iruretagoyena, M. Á. d. l. C. Crespo, M. Alabèrnia-Segura, B. Spanlang, and G. Feixas. An experimental study of a virtual reality counselling paradigm using embodied self-dialogue. *Scientific reports*, 9(1):10903, 2019. 1, 3

[30] C. Tacca, B. Kerr, and E. Friis. The development of a common factors based virtual reality therapy system for remote psychotherapy applications. In *2022 IEEE conference on virtual reality and 3D user interfaces abstracts and workshops (VRW)*, pp. 454–458. IEEE, 2022. 3

[31] B. G. Witmer and M. J. Singer. Measuring presence in virtual environments: A presence questionnaire. *Presence*, 7(3):225–240, 1998. 2